# Raman scattering in transition-metal dichalcogenides MTe$_2$ (M = Mo, W)

Xiaoli Ma,[1] Pengjie Guo,[1] Changjing Yi,[2] Qiaohe Yu,[1] Anmin Zhang,[1] Jianting Ji,[1] Yong Tian,[1] Feng Jin,[1] Yiyan Wang,[1] Kai Liu,[1] Tianlong Xia,[1] Youguo Shi,[2] Qingming Zhang[1,*]

[1]*Department of Physics, Beijing Key Laboratory of Opto-Electronic Functional Materials and Micro-nano Devices, Renmin University of China, Beijing 100872, P. R. China*

[2]*Beijing National Laboratory for Condensed Matter Physics, Institute of Physics, Chinese Academy of Sciences, Beijing 100190, P. R. China*

**Abstract**

We performed comparable polarized Raman scattering studies of MoTe$_2$ and WTe$_2$. By rotating crystals to tune the angle between the principal axis of the crystals and the polarization of the incident/scattered light, we obtained the angle dependence of the intensities for all the observed modes, which is perfectly consistent with careful symmetry analysis. Combining these results with first-principles calculations, we clearly identified the observed phonon modes in the different phases of both crystals. Fifteen Raman-active phonon modes (10A$_g$+5B$_g$) in the high-symmetry phase 1T'-MoTe$_2$ (300 K) were well assigned, and all the symmetry-allowed Raman modes (11A$_1$+6A$_2$) in the low-symmetry phase T$_d$-MoTe$_2$ (10 K) and 12 Raman phonons (8A$_1$+4A$_2$) in T$_d$-WTe$_2$ were observed and identified. The present work provides basic information about the lattice dynamics in transition-metal dichalcogenides and may shed some light on the understanding of the extremely large magnetoresistance (MR) in this class of materials.

**Introduction**

Both MoTe$_2$ and WTe$_2$ are non-magnetic thermoelectric semi-metals [1, 2] discovered in the mid-twentieth century. They have recently attracted a great deal of interest in the fields of condensed matter physics and materials science because of the extremely large magnetoresistance (MR) found in this class of non-magnetic compounds [3-10]. In T$_d$-WTe$_2$, for example, a huge MR at low temperatures is observed when a magnetic field and current are applied along the c-axis and the a-axis, respectively. At 0.53 K, the MR under a 60 T field can reach $1.3 \times 10^7$% with no saturation trend. Ali et al. found that the electron pocket is essentially the same size as the hole pocket [3] at low temperatures in T$_d$-WTe$_2$. This has been experimentally verified using a variety of methods, including angle-resolved photoemission spectroscopy (ARPES) [11-13], quantum oscillation [14, 15] and Hall effect [16] experiments. Experiments performed under high pressure have indicated that the difference between electron and hole pockets increases with increasing pressure, at the same time MR is gradually suppressed [14]. Therefore, the origin of the huge MR may be attributable to the perfect compensation of electrons and hole pockets [3, 11, 14, 15, 17]. In fact, the mechanism of the extremely large MR remains an open question. In addition, superconducting transition has been observed in Td-MoTe$_2$ and Td-WTe$_2$ under pressure [18-20]. For Td-WTe$_2$, T$_c$ can reach 7 K under a pressure of 16.8 GPa [18], and T$_c$ goes up to 8.2 K at P = 11.7 GPa in Td-MoTe$_2$ [20]. Interestingly, very recent studies show that Td-MoTe$_2$ and Td-WTe$_2$ are Type-II Weyl semimetals with a layered structure [21, 22], whose Weyl point exists at the interface of electron and hole pockets. In fact, most other layered transition-metal dichalcogenides (TMDs), including MoS$_2$ [23-27] and WSe$_2$ [28-31], are semiconductors, which is directly related to their different structures.

Generally, TMDs can assume various configurations, including the 2H, 1T, 1T′, and T$_d$ structures. MoTe$_2$ can exist in the 2H (hexagonal, space group *P*6$_3$/*mmc*), 1T′ (monoclinic, space group *P*2$_1$/*m*), or T$_d$ structure (orthorhombic structure, space group *Pnm*2$_1$) [32]. The 1T′ and T$_d$ structures are quite similar, and both of them have a half-metallic state [10, 22]. The symmetry change between the two structures originates from dislocations between stacked layers [22, 33]. With decreasing temperature, a structural transition occurs in MoTe$_2$ from the monoclinic 1T′ phase to the orthogonal T$_d$ phase at ~270 K. The T$_d$ structure of MoTe$_2$ is isomorphic with the T$_d$ structure of WTe$_2$, and both of them exhibit a huge MR effect. WTe$_2$, a layered compound with a half-metallic state [1, 3, 35, 36], usually has a stable T$_d$ structure (crystallized into an orthogonal structure of distorted octahedra at atmospheric pressure, space group *Pnm*2$_1$) [34]. By contrast, many other TMDs are usually semiconductors with 2H or 1T structures. MoTe$_2$/WTe$_2$ and other TMD materials have lately attracted widespread attention because of their rich physics, such as their huge MR effect, superconducting transition under pressure and Weyl semi-metal states, among other phenomena.

Raman scattering is among the most conventional and fundamental techniques for studying TMDCs. It can determine the structure and layer number in an easy and nondestructive manner [37, 38]. Studies of the lattice dynamics of these materials lay the foundation for exploring possible lattice structures and the influence of phonons on electronic energy bands. It is provides an important basis for the study of MoTe$_2$ and WTe$_2$ heterojunctions and their doping and phase transitions [2, 39-41].

In this work, we conducted a systematic Raman study of MoTe$_2$ and WTe$_2$. Fifteen Raman-active modes (10A$_g$+5B$_g$) were measured in 1T′-MoTe$_2$. All 17 A$_1$/A$_2$ Raman modes (11A$_1$+6A$_2$) in T$_d$-MoTe$_2$ and 12 Raman modes (8A$_1$+4A$_2$) in T$_d$-WTe$_2$ were observed. To resolve the observed phonon modes in MoTe$_2$ and WTe$_2$, we performed polarized Raman measurements. The mode intensity modulations induced by rotating the samples confirmed their symmetries, which are in good agreement with strict symmetry analysis. We further conducted first-principles calculations. Combining symmetry analysis with first-principles calculations allowed us to clearly identify all the observed modes. The corresponding vibration patterns were also given.

**Experimental method**

Single crystals of MTe$_2$ (M = Mo, W) were grown by a flux method, as described in the literature [14]. The polarized and angular Raman spectra were collected using a HR800 spectrometer (Jobin Yvon) equipped with liquid-nitrogen-cooled CCD and volume Bragg gratings, for which micro-Raman backscattering configuration was adopted. A 633 nm laser was used, with a spot size of ~5 μm focused on the sample surface. The laser power was maintained at approximately 1.4 mW to avoid overheating during measurements.

X/Y in this paper corresponds to the a/b-axis of crystalline MTe$_2$ (M = Mo, W). X′/Y′ is a 45° angle with respect to X/Y. Z is perpendicular to the XY plane. The angle dependence of the mode intensity was measured by rotating the crystals while the polarizations of the incident and scattered light were fixed.

**Calculation method**

To determine the phonon modes of WTe$_2$ and MoTe$_2$, we conducted first-principles electronic structure calculations using the projector-augmented wave method [42], as implemented in the VASP package [43]. For the exchange-correlation potential, the generalized gradient approximation (GGA) of Perdew-Burke-Ernzerh [44] was adopted. To describe the van der Waals (VDW) interaction in layered systems not included in conventional density functional theory, the vdW-optB86b function [45] was chosen. The kinetic energy cutoff of the plane-wave basis was set to 300 eV. The simulations were performed with an orthorhombic supercell containing 4 W atoms and 8 Te atoms. A 20 × 10 × 8 k-point mesh was employed for the Brillouin zone sampling. Gaussian smearing with a width of 0.01 eV was used around the Fermi surface. In the structure optimization, both the cell parameters and internal atomic positions were allowed to relax until all forces were smaller than 0.005 eV/Å. The calculated lattice parameters were in good agreement with the experimental values (error less than 1.7%) [33]. After the equilibrium structure was obtained, the vibrational frequencies and polarization vectors at the Brillouin zone center were calculated using the dynamic matrix method.

**Results and discussion**

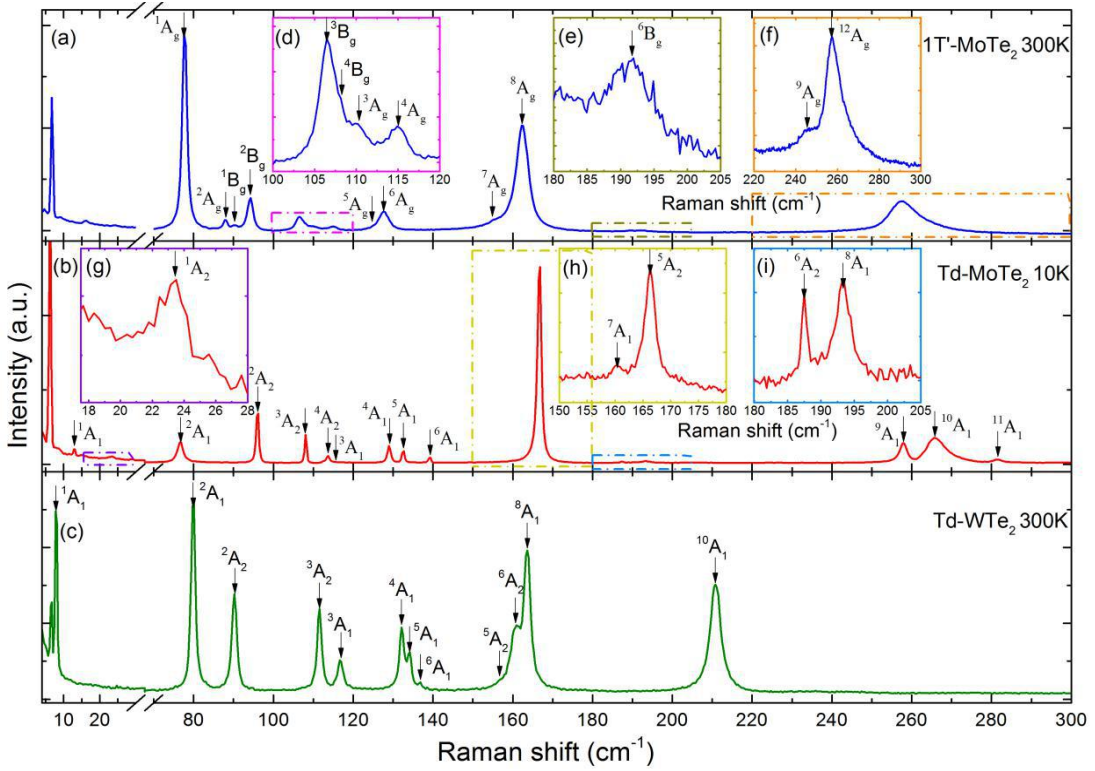

FIG. 1. (a) Raman spectrum of 1T′-MoTe$_2$ measured at 300 K. (d), (e) and (f) are zoomed in to show the modes with weak intensities in (a). (b) Raman spectrum of T$_d$-MoTe$_2$ measured at 10 K. (g), (h) and (i) are zoomed in to show the modes with weak intensities in (b). (c) Raman spectrum of T$_d$-WTe$_2$ measured at 10 K.

Fig. 1(a) shows the Raman spectrum of 1T′-MoTe$_2$ collected at 300 K. 1T′-MoTe$_2$ has a monoclinic structure with space group $P2_1/m(C_{2h})$ [32, 34], which allows 18 Raman-active phonon modes (12 A$_g$ modes and 6 B$_g$ modes). These modes should be visible when the incident light vertically strikes the ab plane. In fact, we observed 15 Raman-active phonon modes. The other three modes are invisible, perhaps due to weak signals. The signal at the lowest frequencies arises from the laser line.

Fig. 1(b), (c) shows the spectrum of T$_d$-MoTe$_2$ measured at 10 K and the spectrum of T$_d$-WTe$_2$ measured at 300 K, respectively. T$_d$-MoTe$_2$ and T$_d$-WTe$_2$ have the same orthorhombic structure with space group $Pnm2_1(C^7_{2v})$ [46]. Symmetry analysis indicates that there should be 33 Raman-active phonon modes

[11$A_1$+6$A_2$+5$B_1$+11$B_2$] in the low-symmetry phase. Our measurements were conducted on the ab plane, which allows 17 Raman modes (11$A_1$ + 6$A_2$). All 17 modes were observed in the spectrum of $T_d$-MoTe$_2$. By comparison, only 12 Raman modes were observed in the spectrum of $T_d$-WTe$_2$ under the same experimental conditions. The reason for this difference may be weak signals of the other five phonon modes. As in the spectrum of 1T′-MoTe$_2$ (Fig. 1(a)), the peak at the lowest frequencies in the spectrum of $T_d$-MoTe$_2$ (Fig. 1(b)) originate from the excitation laser spectrum; by contrast, the peaks at the lowest frequencies in the spectrum of $T_d$-WTe$_2$ (Fig. 1(c)) actually contain a phonon mode in addition to the laser line. To verify the symmetry of the observed modes, we collected the corresponding polarized Raman spectra.

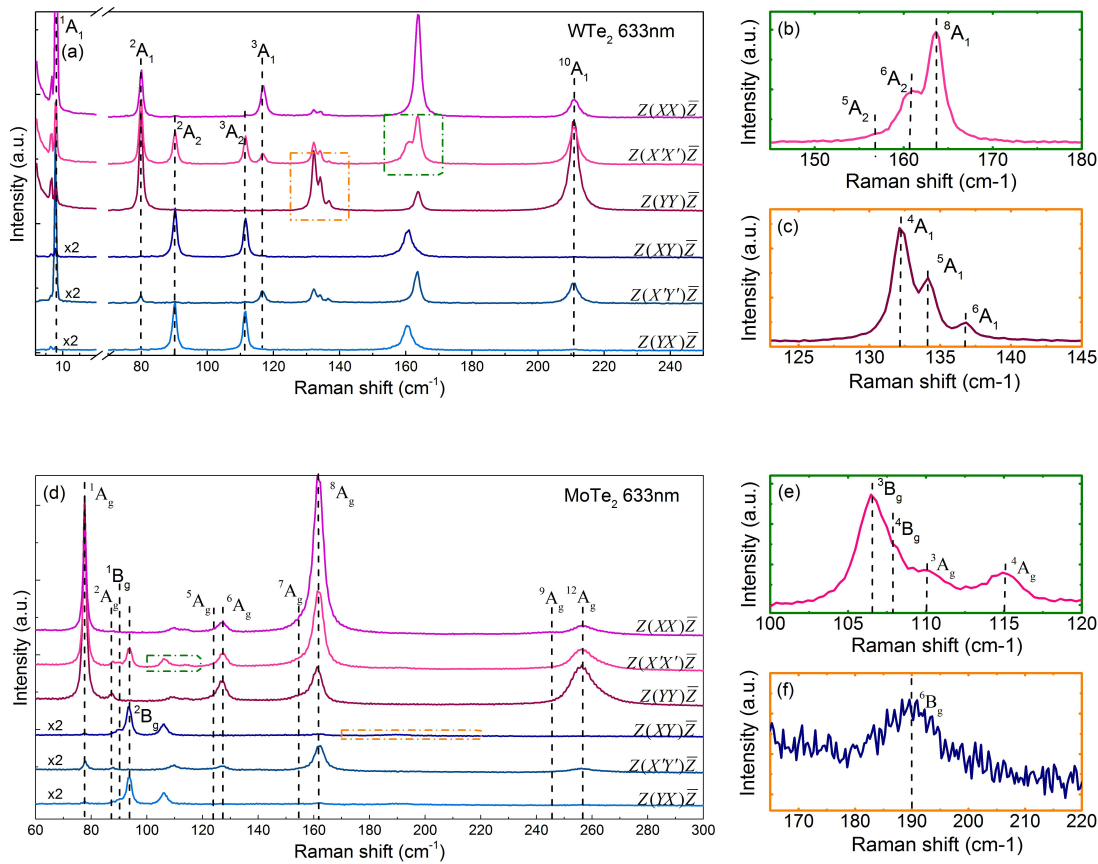

FIG. 2. (a) Polarized Raman spectrum of $T_d$-WTe$_2$ collected at room temperature using a 633 nm laser. (b) and (c) Zoomed-in views of the green/orange-dotted box in Fig. 2(a). (d) Polarized Raman spectrum of 1T′-MoTe$_2$ collected at room temperature using a 633 nm laser. (e) and (f) Zoomed-in views of the green/orange-dotted box in Fig. 2(d).

Fig. 2 shows the polarized Raman spectra of $T_d$-$WTe_2$ and $1T'$-$MoTe_2$ at room temperature under different polarization configurations. Totals of 12 and 15 Raman modes can be seen in the spectra of $T_d$-$WTe_2$ and $1T'$-$MoTe_2$, respectively. Comparison of the spectra of the two similar structures demonstrates that the subtle structural difference is clearly distinguished by Raman scattering.

Bulk $T_d$-$WTe_2$ has an orthogonal structure with the space group $Pnm2_1(C^7_{2v})$ [46], in which layers are stacked along the c-axis. Generally, Raman intensities are determined by Raman tensors. The Raman tensors for $WTe_2$ are as follows:

$$A_1 = \begin{pmatrix} a & 0 & 0 \\ 0 & b & 0 \\ 0 & 0 & c \end{pmatrix}, \quad A_2 = \begin{pmatrix} 0 & d & 0 \\ d & 0 & 0 \\ 0 & 0 & 0 \end{pmatrix}, \quad B_1 = \begin{pmatrix} 0 & 0 & e \\ 0 & 0 & 0 \\ e & 0 & 0 \end{pmatrix}, \quad B_2 = \begin{pmatrix} 0 & 0 & 0 \\ 0 & 0 & f \\ 0 & f & 0 \end{pmatrix}.$$

Symmetry analysis indicates that only the $A_1$ and $A_2$ modes are visible in our case because our measurements were conducted on the ab plane. According to the tensors of the $A_1$ and $A_2$ modes, all 12 Raman modes can be seen in the channel of $Z(X'X')\bar{Z}$ ($e_i // e_s, \theta = 45°$) or $Z(X'Y')\bar{Z}$ ($e_i \perp e_s, \theta = 45°$); by contrast, in $Z(XX)\bar{Z}$ ($e_i // e_s, \theta = 0°$) or $Z(YY)\bar{Z}$ ($e_i // e_s, \theta = 90°$), only 8 $A_1$ modes should be observed. Experimentally, these modes are located at 8.0 cm$^{-1}$ ($^1A_1$), 80.1 cm$^{-1}$ ($^2A_1$), 117.7 cm$^{-1}$ ($^3A_1$), 132.9 cm$^{-1}$ ($^4A_1$), 134.8 cm$^{-1}$ ($^5A_1$), 138.4 cm$^{-1}$ ($^6A_1$), 164.5 cm$^{-1}$ ($^8A_1$), and 212.0 cm$^{-1}$ ($^9A_1$). In the case of $Z(XY)\bar{Z}$ ($e_i \perp e_s, \theta = 0°$) or $Z(YX)\bar{Z}$ ($e_i \perp e_s, \theta = 90°$), only four $A_2$ modes are visible, located at 90.6 cm$^{-1}$ ($^2A_2$), 112.1 cm$^{-1}$ ($^3A_2$), 156.1 cm$^{-1}$ ($^5A_2$), and 161.2 cm$^{-1}$ ($^6A_2$). Through careful polarized Raman measurements, we can easily distinguish the $A_1$ and $A_2$ modes.

Bulk $1T'$-$MoTe_2$ has a monoclinic structure with the space group $P2_1/m(C_{2h})$ [32, 34]. Its Raman tensors are written as follows:

$$A_g = \begin{pmatrix} b & 0 & d \\ 0 & c & 0 \\ d & 0 & a \end{pmatrix}, \quad B_g = \begin{pmatrix} 0 & f & 0 \\ f & 0 & e \\ 0 & e & 0 \end{pmatrix}.$$

Symmetry analysis tells us that the $A_g$ and $B_g$ modes are visible when the measurements are conducted on the ab plane. The tensors of the $A_g$ and $B_g$ modes allow 15 Raman modes in $Z(X'X')\bar{Z}$ ($e_i // e_s, \theta = 45°$) or $Z(X'Y')\bar{Z}$ ($e_i \perp e_s, \theta = 45°$), and all of them were observed in this work. Only 10 $A_g$ modes are observed in $Z(XX)\bar{Z}$ ($e_i // e_s, \theta = 0°$) or $Z(YY)\bar{Z}$ ($e_i // e_s, \theta = 90°$). Experimentally, these modes correspond to 77.0 cm$^{-1}$ ($^1A_g$), 86.3 cm$^{-1}$ ($^2A_g$), 110.7 cm$^{-1}$ ($^3A_g$), 114.5 cm$^{-1}$ ($^4A_g$), 127.8 cm$^{-1}$ ($^5A_g$), 129.0 cm$^{-1}$ ($^6A_g$), 155.4 cm$^{-1}$ ($^7A_g$), 163.3 cm$^{-1}$ ($^8A_g$), 245.0 cm$^{-1}$ ($^9A_g$), and 256.3 cm$^{-1}$ ($^{12}A_g$). In $Z(XY)\bar{Z}$ ($e_i \perp e_s, \theta = 0°$) or $Z(YX)\bar{Z}$ ($e_i \perp e_s, \theta = 90°$), only five $B_g$ modes should be observed, corresponding to 91.2 cm$^{-1}$ ($^1B_g$), 94.4 cm$^{-1}$ ($^2B_g$), 106.1 cm$^{-1}$ ($^3B_g$), 107.0 cm$^{-1}$ ($^4B_g$), and 191.2 cm$^{-1}$ ($^6B_g$). Thus, the $A_g$ or $B_g$ modes can be distinguished through different polarization configurations.

Fig. 3(a,b)/(g,h) show the angle dependence of the Raman intensities for $T_d$-WTe$_2$/1T′-MoTe$_2$ in both parallel and cross channels, demonstrating that the Raman intensities exhibit a regular change as the crystals are rotated. The intensity variations are illustratively summarized in Fig. 3(c-f)/(i-l). Here, $e_i$ and $e_s$ are the polarizations of the incident and scattered light, respectively. The angle between the polarizations of the incident light and the principal axis (axis a) is defined as $\theta$.

According to the aforementioned Raman tensors, the Raman intensities for different symmetries in $T_d$-WTe$_2$ can be written as follows:

$$e_i // e_s: \quad A_1: \quad I = \left[b + (a-b)\cos^2\theta\right]^2 \qquad (1)$$

$$A_2: \quad I = (d\sin 2\theta)^2 \qquad (2)$$

$$e_i \perp e_s: \quad A_1: \quad I = \left(\frac{b-a}{2}\sin 2\theta\right)^2 \qquad (3)$$

$$A_2: \quad I = (d\cos 2\theta)^2 \qquad (4)$$

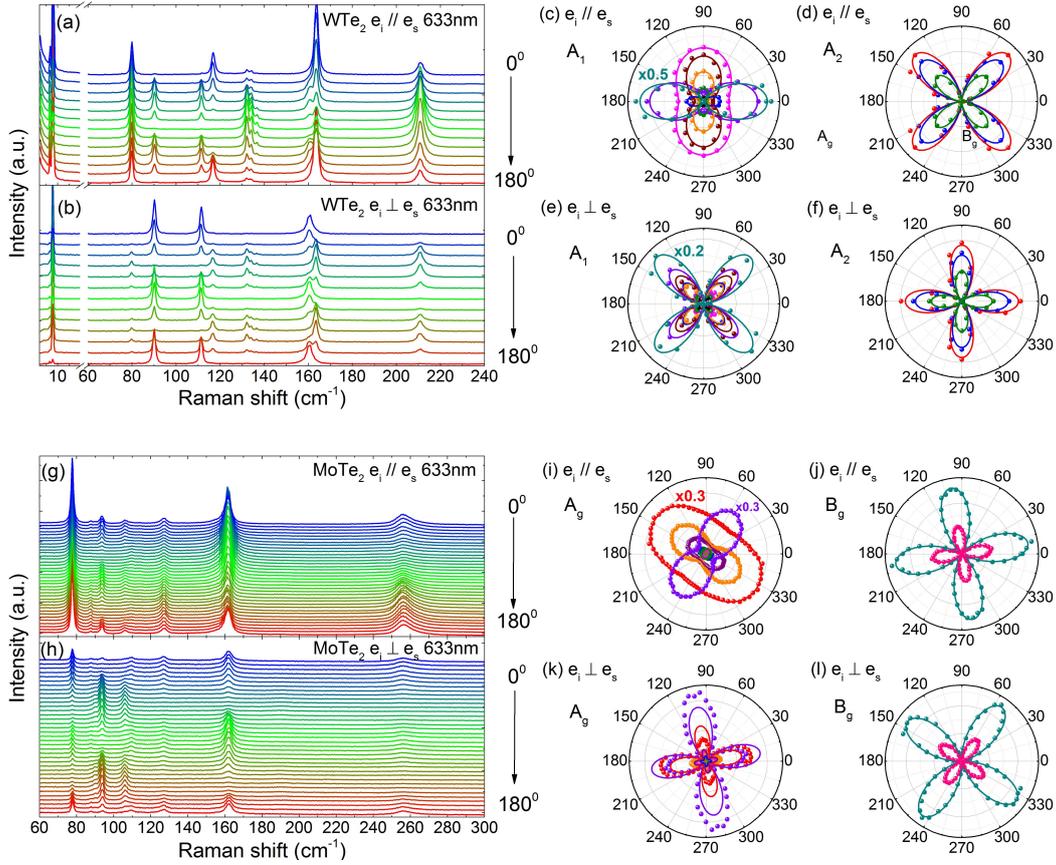

FIG. 3. (a) and (b) Angle dependence of $T_d$-WTe$_2$ in parallel and cross-polarization configurations. (g) and (h) Angle dependence of 1T′-MoTe$_2$ in parallel and cross-polarization configurations. (c)-(f) Intensity plots with respect to rotation angle for the $A_1/A_2$ modes of $T_d$-WTe$_2$. (i)-(l) Intensity plots with respect to rotation angle for the $A_g/B_g$ modes of 1T′-MoTe$_2$.

Similarly, the Raman intensities for the different symmetries in 1T′-MoTe$_2$ have the following forms:

$$e_i // e_s: \quad A_g: \quad I = \left[c+(b-c)\cos^2\theta\right]^2 \quad (5)$$

$$B_g: \quad I = (f\sin 2\theta)^2 \quad (6)$$

$$e_i \perp e_s: \quad A_g: \quad I = \left(\frac{c-b}{2}\sin 2\theta\right)^2 \quad (7)$$

$$B_g: \quad I = (f\cos 2\theta)^2 \quad (8)$$

Although the two structures are similar, polarized Raman measurements can reveal subtle dependencies. The angle dependence for the $A_1(A_g)$ modes in the parallel channel exhibits twofold symmetry, whereas the angle dependence for the $A_2(B_g)$ modes exhibits fourfold symmetry. The angle dependence for the $A_1/A_2$ and $A_g/B_g$ modes in the cross channel exhibits fourfold symmetry; however, the maximum intensities appear at different angles depending on the mode symmetries. Notably, the non-zero starting angle of 1T′-MoTe$_2$ has been taken into account. Fig. 3 (c)-(f), (i)-(l) indicates that the detailed symmetry analysis shows good consistency with the experimental results. This consistency indicates that the symmetry identification for each phonon in T$_d$-WTe$_2$ and 1T′-MoTe$_2$ is reasonable.

To assign the observed modes, we conducted first-principles calculations. The experimental and calculated mode frequencies are summarized in Table 1. The table indicates that the experimental values agree well with the calculated data. In total, there are $11A_1+6A_2+5B_1+11B_2$ vibration modes in T$_d$-MoTe$_2$/T$_d$-WTe$_2$. Among them, six $A_2$ modes are Raman active, and the rest are infrared and Raman-active. A total of $12A_g+6B_g+5A_u+10B_u$ vibration modes are allowed in 1T′-MoTe$_2$. Of these modes, 12 $A_g$ modes and 6 $B_g$ modes are Raman-active and the remaining 5 $A_u$ modes and 10 $B_u$ modes are infrared-active.

TABLE I. Comparison of the calculated and experimental optical phonon modes (in cm$^{-1}$) for T$_d$-MoTe$_2$/T$_d$-WTe$_2$ and 1T′-MoTe$_2$. The main atomic motions of the modes are also given (see Fig. 4). *I* and *R* denote infrared and Raman activities, respectively.

| | T$_d$-MoTe$_2$ / T$_d$-WTe$_2$ | | | 1T′-MoTe$_2$ | | | |
|---|---|---|---|---|---|---|---|
| Symmetry | Calculation | Experiment | Activity | Symmetry | Calculation | Experiment | Activity |
| $^1$A$_1$ | 10.3 / 9.5 | 13.0 / 8.0 | I+R | $^1$B$_u$ | 8.8 | | I |
| $^1$A$_2$ | 26.9 / 24.3 | 23.6 | R | $^1$A$_u$ | 26.3 | | I |
| $^1$B$_2$ | 32.1 / 28.2 | | I+R | $^2$B$_u$ | 32.5 | | I |
| $^2$A$_1$ | 77.4 / 76.2 | 76.7 / 80.1 | I+R | $^1$A$_g$ | 76.6 | 77.0 | R |
| $^2$B$_2$ | 85.8 / 86.1 | | I+R | $^2$A$_g$ | 84.8 | 86.3 | R |
| $^1$B$_1$ | 88.5 / 87.1 | | I+R | $^1$B$_g$ | 88.2 | 91.2 | R |
| $^2$A$_2$ | 91.2 / 88.1 | 96.1 / 90.6 | R | $^2$B$_g$ | 91.0 | 94.4 | R |
| $^3$A$_2$ | 104.6 / 109.7 | 108.1 / 112.1 | R | $^3$B$_g$ | 104.0 | 106.1 | R |
| $^2$B$_1$ | 105.3 / 110.0 | | I+R | $^4$B$_g$ | 104.7 | 107.0 | R |
| $^4$A$_2$ | 108.0 / 113.0 | 113.7 | R | $^2$A$_u$ | 107.9 | | I |
| $^3$A$_1$ | 108.7 / 113.2 | 115.5 / 117.7 | I+R | $^3$A$_g$ | 109.4 | 110.7 | R |
| $^3$B$_1$ | 110.6 / 115.2 | | I+R | $^3$A$_u$ | 110.4 | | I |
| $^3$B$_2$ | 113.0 / 117.6 | | I+R | $^4$A$_g$ | 113.6 | 114.5 | R |
| $^4$B$_2$ | 115.1 / 123.0 | | I+R | $^3$B$_u$ | 116.0 | | I |
| $^4$A$_1$ | 122.4 / 128.5 | 129.1 / 132.9 | I+R | $^4$B$_u$ | 123.4 | | I |
| $^5$A$_1$ | 125.1 / 130.6 | 132.6 / 134.8 | I+R | $^5$A$_g$ | 125.5 | 127.8 | R |
| $^5$B$_2$ | 127.3 / 132.3 | | I+R | $^6$A$_g$ | 127.0 | 129.0 | R |
| $^6$B$_2$ | 128.9 / 127.4 | | I+R | $^5$B$_u$ | 129.0 | | I |
| $^6$A$_1$ | 134.2 / 132.0 | 139.2 / 138.4 | I+R | $^6$B$_u$ | 133.9 | | I |
| $^7$B$_2$ | 153.7 / 154.8 | | I+R | $^7$A$_g$ | 153.2 | 155.4 | R |
| $^7$A$_1$ | 158.6 / 158.8 | 160.4 | I+R | $^8$A$_g$ | 158.5 | 163.3 | R |
| $^5$A$_2$ | 175.9 / 151.0 | 166.7 / 156.1 | R | $^4$A$_u$ | 175.5 | | I |
| $^4$B$_1$ | 176.3 / 151.2 | | I+R | $^5$A$_u$ | 176.2 | | I |
| $^6$A$_2$ | 188.4 / 159.5 | 187.5 / 161.2 | R | $^5$B$_g$ | 186.9 | | R |
| $^5$B$_1$ | 188.9 / 161.0 | | I+R | $^6$B$_g$ | 187.9 | 191.2 | R |
| $^8$A$_1$ | 191.2 / 171.9 | 193.3 / 164.5 | I+R | $^7$B$_u$ | 190.8 | | I |
| $^8$B$_2$ | 191.4 / 170.7 | | I+R | $^8$B$_u$ | 191.6 | | I |
| $^9$A$_1$ | 244.9 / 209.1 | 258.1 / 212.0 | I+R | $^9$A$_g$ | 242.0 | 245.0 | R |
| $^9$B$_2$ | 245.6 / 210.6 | | I+R | $^{10}$A$_g$ | 246.8 | | R |
| $^{10}$B$_2$ | 251.6 / 205.8 | | I+R | $^{11}$A$_g$ | 252.4 | | R |
| $^{10}$A$_1$ | 251.6 / 205.6 | 265.4 | I+R | $^{12}$A$_g$ | 254.9 | 256.3 | R |
| $^{11}$B$_2$ | 265.1 / 231.7 | | I+R | $^9$B$_u$ | 266.6 | | I |
| $^{11}$A$_1$ | 265.9 / 231.2 | 280.4 | I+R | $^{10}$B$_u$ | 267.3 | | I |

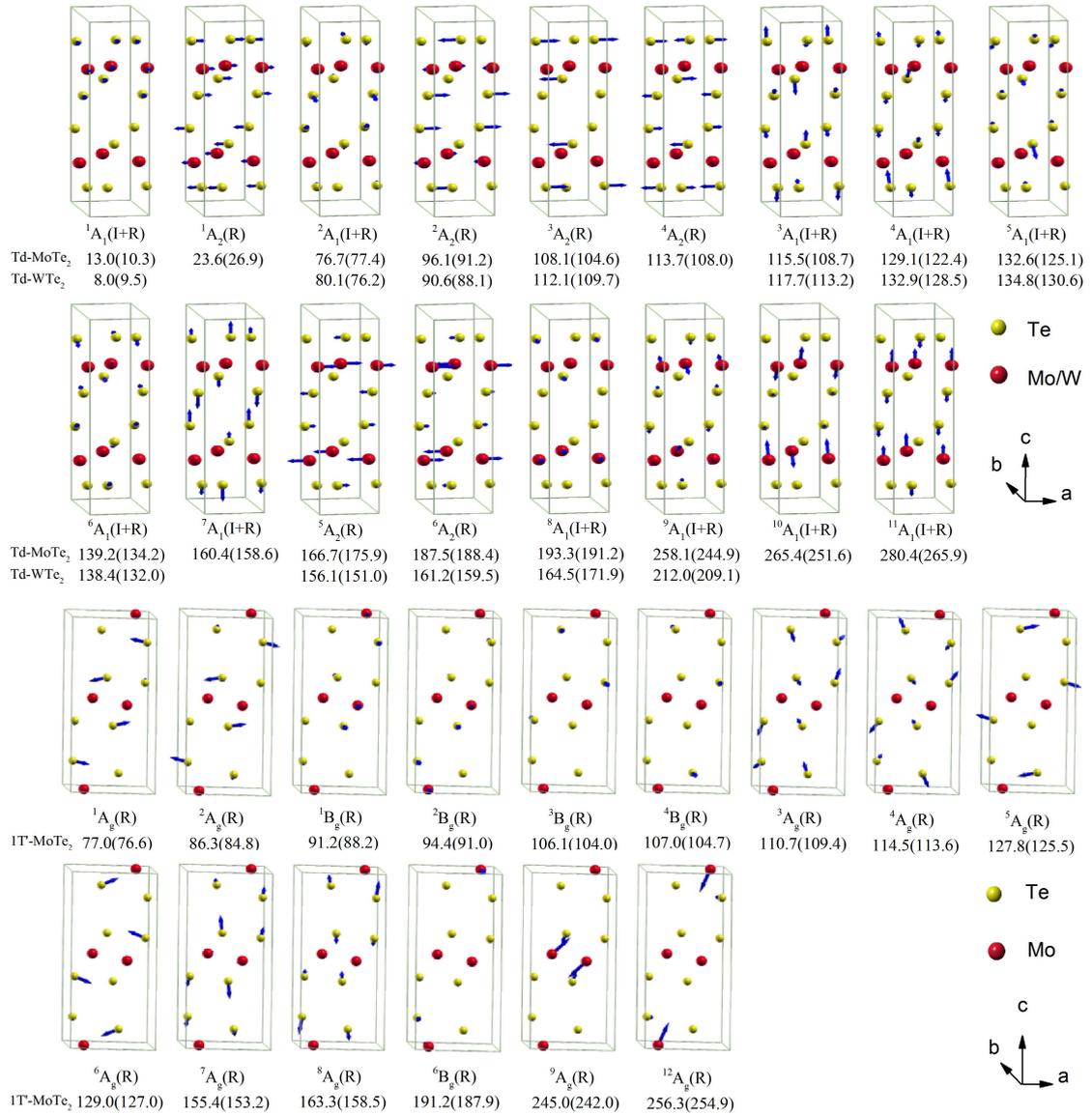

FIG. 4. Vibration patterns of all the Raman modes observed in $T_d$-MoTe$_2$/$T_d$-WTe$_2$ and 1T′-MoTe$_2$. The mode symmetry, optical activity and experimental (calculated) phonon frequency are also listed below each pattern. I/R indicates infrared/Raman activity.

The vibration patterns of the observed phonon modes are illustrated in Fig. 4. $T_d$-WTe$_2$ and $T_d$-MoTe$_2$ have the same structure and identical vibration patterns. The first two rows of Fig. 4 show the vibration patterns of $T_d$-WTe$_2$ and $T_d$-MoTe$_2$. Because the effective mass m of W is larger than that of Mo and because $\omega \propto \sqrt{k/m}$, the phonon frequency $\omega$ of $T_d$-WTe$_2$ is expected to be smaller than that of $T_d$-MoTe$_2$ for the same mode. The calculations indicate that most of the modes follow this expectation. However, a few phonon modes appear anomalous, which may be due to the change in the relevant spring constants. Meanwhile, the experimental data are consistent with the calculations, which supports the previous speculation concerning $\omega$. The last two rows of Fig. 4 show the vibration patterns of 1T′-MoTe$_2$. The phase transition from the 1T′-phase to the T$_d$-phase in MoTe$_2$ is correlated with the stacked misalignment between the layers. A comparison between the interlayer modes of 1T′-MoTe$_2$ (e.g., $^7$A$_g$) and the interlayer modes of T$_d$-MoTe$_2$ (e.g., $^7$A$_1$) reveals that the corresponding vibration patterns show only small changes across the phase transition. This result suggests that the phase transition has little effect on the interlayer modes.

**Conclusion**

This paper gives a comprehensive identification of the vibration modes for 1T′-MoTe$_2$, T$_d$-MoTe$_2$ and T$_d$-WTe$_2$. The polarized Raman spectra of bulk 1T′-MoTe$_2$, T$_d$-MoTe$_2$ and T$_d$-WTe$_2$ were collected, and 15, 17 and 12 Raman-active phonon modes were observed, respectively. The A$_1$ (A$_g$) and A$_2$ (B$_g$) modes were distinguished by careful polarized measurements combined with group-theory analysis. First-principles calculations were conducted to identify the vibration pattern for each mode. Comparison of the vibration patterns of the interlayer modes of 1T′-MoTe$_2$ and T$_d$-MoTe$_2$, revealed that the phase transition has very little influence on the interlayer modes. These results may have important implications for studies of MoTe$_2$ and WTe$_2$ heterojunctions, doping, and phase transitions, among other topics.

**Acknowledgments**

This work was supported by the Ministry of Science and Technology of China (Grant No. 2016YFA0300504) and the NSF of China. Y.G.S was supported by the Strategic Priority Research Program (B) of the Chinese Academy of Sciences (Grant No. XDB07020100). Q.M.Z., A.M.Z., K.L. and T.L.X were supported by the Fundamental Research Funds for the Central Universities and the Research Funds of Renmin University of China (10XNI038, 14XNLF06, 14XNLQ03 and 14XNLQ07). Computational resources were provided by the Physical Laboratory of High

Performance Computing at Renmin University of China. The atomic structures and vibrational displacement patterns were prepared using the XCRYSDEN program [47].